\documentclass[aps,prl,twocolumn,superscriptaddress,showpacs]{revtex4}
\usepackage{graphicx}
\usepackage{dcolumn}
\usepackage{bm}
\usepackage{amsmath}

\begin{document}

\title{
Magnetic-Field-Induced $4f$-Octupole in CeB$_6$ Probed by Resonant X-ray Diffraction
}


\author{Takeshi Matsumura}
\email[]{tmatsu@hiroshima-u.ac.jp}
\affiliation{Department of Quantum Matter, AdSM, Hiroshima University, Higashi-Hiroshima, 739-8530, Japan}
\author{Takumi Yonemura}
\affiliation{Department of Quantum Matter, AdSM, Hiroshima University, Higashi-Hiroshima, 739-8530, Japan}
\author{Keisuke Kunimori}
\affiliation{Department of Quantum Matter, AdSM, Hiroshima University, Higashi-Hiroshima, 739-8530, Japan}
\author{Masafumi Sera}
\affiliation{Department of Quantum Matter, AdSM, Hiroshima University, Higashi-Hiroshima, 739-8530, Japan}
\author{Fumitoshi Iga}
\affiliation{Department of Quantum Matter, AdSM, Hiroshima University, Higashi-Hiroshima, 739-8530, Japan}


\date{\today}

\begin{abstract}
CeB$_6$, a typical $\Gamma_8$-quartet system, exhibits a mysterious antiferroquadrupolar ordered phase in magnetic fields, which is considered as originating from the $T_{xyz}$-type magnetic octupole moment induced by the field. 
By resonant x-ray diffraction in magnetic fields, we have verified that the $T_{xyz}$-type octupole is indeed induced in the $4f$-orbital of Ce with a propagation vector $(\frac{1}{2}, \frac{1}{2}, \frac{1}{2})$, thereby supporting the theory. 
We observed an asymmetric field dependence of the intensity for an electric quadrupole (E2) resonance when the field was reversed, and extracted a field dependence of the octupole by utilizing the interference with an electric dipole (E1) resonance. 
The result is in good agreement with that of the NMR-line splitting, which reflects the transferred hyperfine field at the Boron nucleus from the anisotropic spin distribution of Ce with an $O_{xy}$-type quadrupole. 
The field-reversal method used in the present study opens up the possibility of being widely applied to other multipole ordering systems such as NpO$_2$, Ce$_x$La$_{1-x}$B$_6$, SmRu$_4$P$_{12}$, and so on. 
\end{abstract}

\pacs{61.05.cp 
, 71.27.+a	
, 75.25.+z 
, 75.10.Dg 
}

\maketitle


A rich variety of electronic phases arising from multiple degrees of freedom of $f$ electrons has attracted great interest in recent years.  
In addition to magnetic dipole moment, electric quadrupole and magnetic octupole, etc., behave as independent degrees of freedom in a crystal-field eigenstate with orbital degeneracy. 
Quadrupole orders are frequently realized in localized $f$-electron systems and, more exotically, octupole orders can also take place as in NpO$_2$ and Ce$_x$La$_{1-x}$B$_6$  $(x\le 0.8)$~\cite{Paixao02,Tokunaga06,Kubo04,Mannix05,Kuwahara07}. 
Furthermore, it has recently been recognized that these multipoles sometimes play fundametal roles when $f$ electrons are hybridized with conduction electrons. 
In Pr-based filled skutterudites such as PrRu$_4$P$_{12}$, a $4f$-hexadecapole order is combined with a Fermi-surface nesting, causing a metal-insulator transition~\cite{Takimoto06}. 
In PrOs$_4$Sb$_{12}$, it is suggested that a quadupolar excitation is associated with the heavy-fermion superconductivity~\cite{Kuwahara05}. 
Thus, understanding the physics of multipole moments is of fundamental importance. 
 
One of the difficulties of multipole research is that they are hard to identify as is often expressed as \textit{hidden order paremeter}. 
In most cases, a primary order parameter (OP) is initially inferred indirectly by combining various pieces of information from macroscopic and microscopic methods. 
Then, detailed investigation of a secondary OP by neutron and x-ray diffraction, and of a hyperfine field by NMR, using a single crystal, may provide evidence for the multipole OP~\cite{Paixao02,Tokunaga06}. 
Among these microscopic probes, resonant x-ray diffraction (RXD) has a distinctive ability to directly probe ordered structures of multipole tensors up to rank 4, using an electric quadrupole (E2) resonant process~\cite{Lovesey05}. 
With respect to the observation of an antiferroquadrupole (AFQ) order, there have already been several examples of successful applications of RXD mainly using an electric dipole (E1) resonance, typically for DyB$_2$C$_2$~\cite{Matsumura05}. 
On the other hand, with respect to an antiferrooctupole (AFO) order, there has been only one report on Ce$_{0.7}$La$_{0.3}$B$_{6}$ by Mannix \textit{et al.}~\cite{Mannix05}.  
They successfully detected an E2 signal at zero field, measured the azimuthal-angle dependence, and concluded an AFO order, which was also confirmed by neutron diffraction~\cite{Kuwahara07}. 
However, it was pointed out that the contribution from the $4f$-quadrupole to the E2 signal cannot be ruled out~\cite{Nagao06}. 
Azimuthal dependence only is not sufficient to separate contributions from different rank tensors to an E2 signal. 

In this Letter, we report an effective method that can distinguish between even and odd rank tensors, which will be quite useful in studying octupole orders, especially those induced in magnetic fields. 
Since various kinds of multipoles are induced in magnetic fields and affects the macroscopic properties, it is of fundamental importance to trace what kind of multipole is induced in magnetic fields. 

A compound we study is CeB$_6$, a typical $\Gamma_8$-quartet system with a simple cubic structure. 
The $\Gamma_8$ has 15 degrees of freedom in total, 3 dipoles, 5 quadrupoles, and 7 octupoles~\cite{Shiina97}. 
At zero field, CeB$_6$ exhibits an $O_{xy}$-type AFQ order at $T_{\text{Q}}$=3.3 K followed by an antiferromagnetic (AFM) order at $T_{\text{N}}$=2.3 K~\cite{Effantin85,Nakao01,Yakhou01}. 
In magnetic fields, $T_{\text{Q}}$ exhibits an anomalous increase up to 8.3 K at 15 T~\cite{Effantin85}, whose most important mechanism has been ascribed to an antiferro-type interaction between field-induced octupoles of $T_{xyz}$-type~\cite{Shiina97}. 
Splitting of the Boron-NMR line in the AFQ phase can be a strong evidence for this interpretation~\cite{Sakai97,Shiina98}. 
It is explained by a phenomenological analysis of the hyperfine field at the Boron nucleus in terms of the multipole moments of Ce based on symmetry arguments. 
To be exact, however, we have to mention that direct evidence for the existence of octupole is still lacking. 
As pointed out by Hanzawa, the microscopic mechanism of the NMR splitting is due to the transferred hyperfine field (THF) via the $2p$ and $2s$ conduction electrons, reflecting the anisotropic spin distribution of Ce with an $O_{xy}$-type quadrupole~\cite{Hanzawa00}. 
The splitting can also be explained phenomenologically if one considers that the THF, e.g., for Borons along the $z$-axis, is proportional to $m_{\text{F}}\langle O_{xy} \rangle$ where  $m_{\text{F}}$ is a field-induced uniform magnetization~\cite{Tsuji01}. 
However, $m_{\text{F}}\langle O_{xy} \rangle$ is not identical to $T_{xyz}$ in the sense that $T_{xyz}\!\equiv\! \frac{\sqrt{5}}{3}(J_{x}O_{yz}\!+\!J_{y}O_{zx}\!+\!J_{z}O_{xy})$ represents a complex magnetization distribution where all the three terms are equally induced even for the field along the $z$-axis. 
The field-induced octupole is essentially a quantum mechanical phenomenon and the increase of $T_{\text{Q}}$ in magnetic fields requires consideration of this real $T_{xyz}$ $4f$-octupole. 
Therefore, it is worth verifying whether the $T_{xyz}$-octupole is induced in the Ce $4f$-orbital itself. 

RXD experiment has been performed at Beamline 3A of the Photon Factory in KEK, using a vertical field superconducting magnet on a two-axis diffractometer. 
A sample with a mirror-polished (331) surface was mounted in the cryostat, so that the [001] and [110] axes were in the horizontal scattering plane and the field was along the [$\bar{1}$10] axis. 
The incident photon was $\pi$-polarized and the energy was tuned to the Ce $L_{\text{III}}$ absorption edge. 

\begin{figure}[t]
\begin{center}
\includegraphics[width=7.5cm]{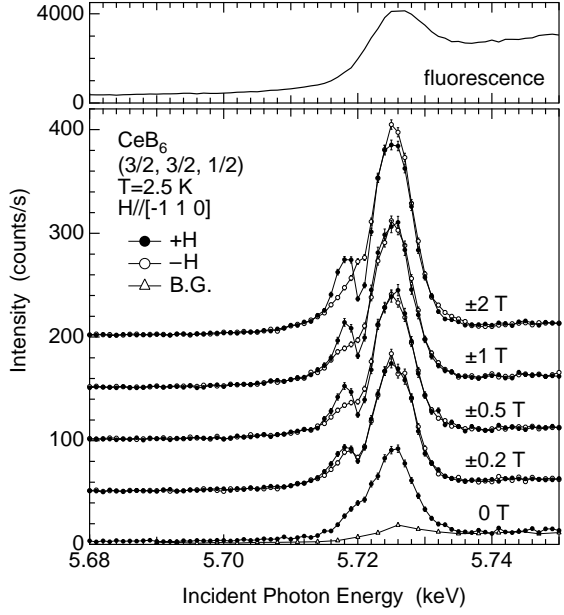}
\end{center}
\caption{top: Fluorescence spectrum of CeB$_{6}$. 
bottom: X-ray energy spectra of the $(\frac{3}{2}\ \frac{3}{2}\ \frac{1}{2})$ superlattice reflection in magnetic fields with reversed directions. 
The triangle shows the background due to the fluorescence. }
\label{fig1}
\end{figure}

Figure~\ref{fig1} shows the energy spectra of the $(\frac{3}{2}\ \frac{3}{2}\ \frac{1}{2})$ superlattice reflection at 2.5 K in the AFQ phase for several magnetic fields with reversed directions. 
We immediately notice that the peaks at 5.724 keV (E1) and at 5.718 keV (E2) become stronger and well resolved for fields in the plus direction, whereas for fields in the minus direction the E2 peak becomes obscure. From this result, we can extract the field dependence of the quadrupole and octupole moments as explained next.

The energy- and field-dependent structure factor for resonant diffraction is generally expressed as 
\begin{align}
F_{\text{reso}}(E,H)=&Z_{\text{E1}}(H)\{f_{\text{E1}}'(E)+if_{\text{E1}}''(E)\} \nonumber \\
+&Z_{\text{E2}}(H)\{f_{\text{E2}}'(E)+if_{\text{E2}}''(E)\} ,
\label{eq:1}
\end{align}
where $Z_{\text{E1}}$ and $Z_{\text{E2}}$ are unit-cell structure factors for E1 and E2 processes, 
which are directly coupled with the atomic tensors $\langle T^{(K)}_{q} \rangle$ of $5d$ and $4f$ orbitals, respectively~\cite{Lovesey05}. 
They are written as 
\begin{align}
Z_{\text{E1}}  = & \sum_{n}e^{i\bm{\kappa}\cdot\bm{R}_n} \sum_{K=0}^{2} \sum_{q} A_K \langle  T^{(K)}_{q} \rangle_{n}^{(5d)} X^{(K)}_{-q} (-1)^q \label{eq:2} \\
Z_{\text{E2}}  = & \sum_{n}e^{i\bm{\kappa}\cdot\bm{R}_n} \sum_{K=0}^{4} \sum_{q} B_K \langle  T^{(K)}_{q} \rangle_{n}^{(4f)} H^{(K)}_{-q} (-1)^q  \label{eq:3} .
\end{align}
Here, $A_K$ and $B_K$ are constant factors for the rank-$K$ terms, $X^{(K)}$ and $H^{(K)}$ are spherical tensors of the x-ray beam determined by the diffraction geometry, $\bm{R}_n$ is a position vector of the $n$th Ce ion in a unit cell, and $\bm{\kappa}$ is a scattering vector. $\langle  T^{(K)}_{q} \rangle$ varies with the applied field. 
The energy dependent term in Eq.~(\ref{eq:1}) can be written as $f(E)=1/(E-\Delta+i\Gamma/2)$ ($\Delta=\Delta_{\text{E1}}$ or $\Delta_{\text{E2}}$) when a resonance can be modeled by a single oscillator. 
However, we leave it here as $f'(E)+if''(E)$ because the actual form is not such simple~\cite{Nagao06}. 

The asymmetry with respect to the field reversal can be understood by considering the following two effects. 
The first is that the E1 and E2 terms interfere, i.e., the intensity is proportional to $|Z_{\text{E1}}f_{\text{E1}}+Z_{\text{E2}}f_{\text{E2}}|^2$ 
and not to $|Z_{\text{E1}}f_{\text{E1}}|^2+|Z_{\text{E2}}f_{\text{E2}}|^2$. 
The second is that the odd rank tensor (magnetic dipole and octupole) reverses its sign with the field reversal, 
whereas the even rank tensor (electric quadrupole and hexadecapole) do not change sign. 
That is, the even(odd) rank terms in $Z$ are symmetric(asymmetric) with respect to the field reversal. 
To analyze the symmetry and asymmetry of the intensity, we write the $Z$ factor in Eq.~(\ref{eq:1}) as $Z^{\text{s}}+iZ^{\text{a}}$, 
where $Z^{\text{s}}$($Z^{\text{a}}$) represents the symmetric(asymmetric) part corresponding to the even(odd) rank term. 
It is noted that the odd rank term is imaginary. 
The energy and field dependent intensity $I(E,H)$ can be calculated by $|F_{\text{reso}}(E,H)|^2$, 
and the symmetric and asymmetric part of the intensity, $I^{\text{s}}(E,H)$ and $I^{\text{a}}(E,H)$, are obtained by $\{I(E,H)+I(E,-H)\}/2$ and 
$\{I(E,H)-I(E,-H)\}/2$, respectively. They are expressed as 
\begin{align}
I^{\text{s}}(E,H)=&\{(Z_{\text{E1}}^{\text{s}})^2+(Z_{\text{E1}}^{\text{a}})^2\}\{(f_{\text{E1}}')^2+(f_{\text{E1}}'')^2\}  \nonumber \\
+&\{(Z_{\text{E2}}^{\text{s}})^{\;2}+(Z_{\text{E2}}^{\text{a}})^2\}\{(f_{\text{E2}}')^2+(f_{\text{E2}}'')^2\} \nonumber \\
+&2(Z_{\text{E1}}^{\text{s}} Z_{\text{E2}}^{\text{s}} +Z_{\text{E1}}^{\text{a}} Z_{\text{E2}}^{\text{a}}) \text{Re} \{f_{\text{E1}}^{*}f^{ }_{\text{E2}}\} \label{eq:4} \\
I^{\text{a}}(E,H)=&2(Z_{\text{E1}}^{\text{a}} Z_{\text{E2}}^{\text{s}} - Z_{\text{E1}}^{\text{s}} Z_{\text{E2}}^{\text{a}}) \text{Im} \{f_{\text{E1}}^{*}f^{ }_{\text{E2}}\}  .\label{eq:5}
\end{align}

\begin{figure}[tb]
\begin{center}
\includegraphics[width=7.5cm]{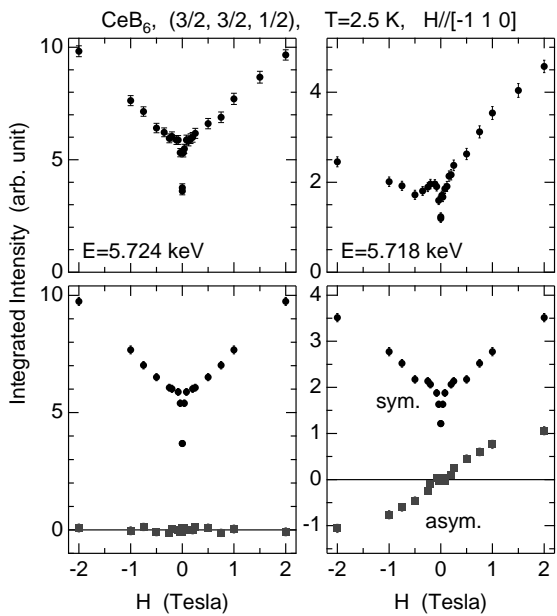}
\end{center}
\caption{top: Magnetic-field dependence of the integrated intensity of the $(\frac{3}{2}\ \frac{3}{2}\ \frac{1}{2})$ resonant Bragg diffraction measured at $E=5.724$ keV (left) and 5.718 keV (right), corresponding to the E1 and E2 processes, respectively. 
bottom: Symmetric (circles) and asymmetric (squares) components of the intensity for the respective processes. 
}
\label{fig2}
\end{figure}

In Fig.~\ref{fig2}, we show the field dependence of the integrated intensity for a rocking scan at each resonance energy. 
The symmetric and asymmetric components deduced from the raw data are shown in the bottom figures. 
At $E=\Delta_{\text{E1}}$ the asymmetric intensity is negligibly small, whereas it clearly exists at $E=\Delta_{\text{E2}}$. 
From these data and Eqs.~(\ref{eq:4}) and (\ref{eq:5}), the field dependence of the multipole tensors can be extracted. 
Of course, to determine all the parameters in general, we need information from azimuthal-angle dependence, polarization analysis, model calculation, and also from other experimental results. 
In the present case of CeB$_6$ for $H\parallel [\bar{1}\ 1\ 0]$, however, some factors can be neglected and the situation become quite simple and suited for a demonstration. 

The sharp anomaly in intensity around 0.1~T corresponds to the one reported in \cite{Brandt85}. 
This is a phase transition from the $\langle O_{xy} \rangle$-OP at zero field, with $\langle O_{yz} \rangle$ and $\langle O_{zx} \rangle$ domains equally populated, to the $\langle \alpha O_{yz} + \beta O_{zx} + \gamma O_{xy} \rangle$-OP, where $(\alpha,\beta,\gamma)$ is the unit vector of the field direction. 
This has also been observed by non-resonant x-ray diffraction~\cite{Tanaka05}. 
Although this is also an important nature of the AFQ phase of CeB$_6$, we do not dealt with it because it is outside the subject of this Letter.

\begin{figure}[tb]
\begin{center}
\includegraphics[width=7.5cm]{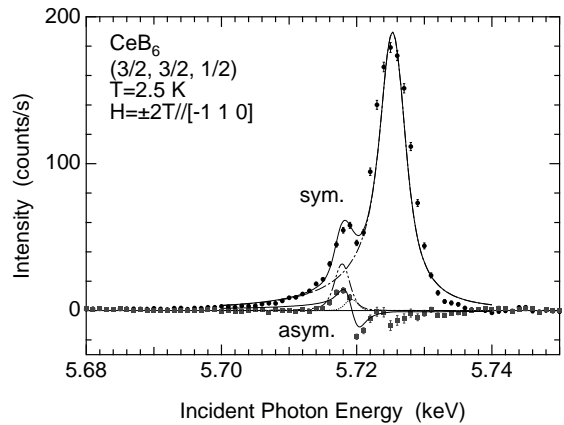}
\end{center}
\caption{X-ray energy dependence of the symmetric and asymmetric components deduced from the data for $\pm$2 T. 
The solid lines are the fits to the data with Lorentzian components. The single dotted, double dotted, and dotted lines represent 1st, 2nd, and 3rd term in Eq.~(\ref{eq:4}), respectively.
}
\label{fig3}
\end{figure}

Figure~\ref{fig3} shows the energy spectra of $I^{\text{s}}(E)$ and $I^{\text{a}}(E)$ deduced from the data for $\pm$2 T.  
We observe that $I^{\text{s}}(E=\Delta_{\text{E1}})$ is dominated by the first term in Eq.~(\ref{eq:4}). 
In fitting $I^{\text{s}}(E)$ and $I^{\text{a}}(E)$, every term in Eqs.~(\ref{eq:4}) and (\ref{eq:5}) was assumed as a Lorentizan, where the real and imaginary parts of $f^{*}_{\text{E1}}f^{ }_{\text{E2}}$ are connected by the Kramers-Kronig relation. 
Absorption effect and a Gaussian resolution of 2 eV were also taken into account in the fit. 
Although there are two contributions from $Z_{\text{E1}}^{\text{s}}$ and $Z_{\text{E1}}^{\text{a}}$ to $I^{\text{s}}(E=\Delta_{\text{E1}})$, $Z_{\text{E1}}^{\text{a}}$, reflecting the field-induced AFM dipole, can be neglected here. 
This is justified by the variation of $I(E=\Delta_{\text{E1}})$ as measured by rotating the crystal around the [331] axis.  
The result can be perfectly explained by considering only the AFQ-OP of $\langle \alpha O_{yz} + \beta O_{zx} + \gamma O_{xy} \rangle$, indicating negligible contribution from the induced AFM. 
In addition, below $T_{\text{N}}$, we could not detect any signal at superlattice spots of the AFM order such as $(\frac{5}{4}\ \frac{5}{4}\ \frac{1}{2})$, probably because it was too small. 
The dipole moment in the AFM phase estimated by neutron diffraction is $\sim$0.28 $\mu_{\text{B}}$~\cite{Effantin85}, whereas that for the induced AFM in the AFQ phase is $\sim$0.05 $\mu_{\text{B}}$ at $H$=2 T~\cite{Tsuji01}. 
That is, the dipole is not the main polarization of the $4f$ shell giving rise to the resonant signal.

\begin{figure}[tb]
\begin{center}
\includegraphics[width=7.5cm]{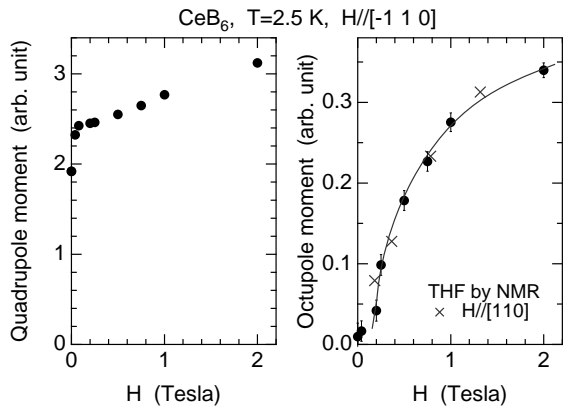}
\end{center}
\caption{Magnetic-field dependences of the AFQ and AFO moments deduced from the symmetric E1 and asymmetric E2 intensities in Fig.~\ref{fig2}. The solid line is a guide for the eye. The crosses represent the THF as deduced from NMR. 
}
\label{fig4}
\end{figure}

By taking the square root of $I^{\text{s}}(E=\Delta_{\text{E1}})$, $Z_{\text{E1}}^{\text{s}}$, reflecting the AFQ moment of the $5d$ orbital, is obtained. 
This is proportional to that of the $4f$ orbital, which is $\langle O_{yz} - O_{zx} \rangle$ for $H\parallel [\bar{1}\ 1\ 0]$ from the structure-factor calculation. 
Next, since $Z_{\text{E1}}^{\text{a}}$ in Eq.~(\ref{eq:5}) can be neglected, we can deduce $Z_{\text{E2}}^{\text{a}}$ by dividing $I^{\text{a}}(E=\Delta_{\text{E2}})$ by $Z_{\text{E1}}^{\text{s}}$. 
As described above, the dipole contribution to $Z_{\text{E2}}^{\text{a}}$ may also be neglected, the obtained result is considered as reflecting only the $4f$-octupole. From the structure-factor calculation, $Z_{\text{E2}}^{\text{a}}$ is proportional to $\langle T_{xyz}+0.02T_{1u}^{z}\rangle$ for $H\parallel [\bar{1}\ 1\ 0]$, where $\langle T_{xyz} \rangle$ is dominant.  
These results are plotted in Fig.~\ref{fig4}. 
$Z_{\text{E2}}^{\text{s}}$, reflecting the $4f$-quadrupole and hexadecapole with the same symmetry of $T_{2g}^{yz}-T_{2g}^{zx}$, can also be deduced after some data treatments, but this results in the same field dependence as that of $Z_{\text{E1}}^{\text{s}}$ as expected. 

In Fig.~\ref{fig4}, the field dependence of the $4f$-octupole shows a good agreement with that of THF at the Boron site as deduced from NMR~\cite{Tsuji01,TakigawaPhD}. 
In addition, it exhibits a convex dependence like a Brillouin function. 
This is quite a contrast to the concave field dependence of the induced AFM as measured by neutron diffraction~\cite{RossatMignod87}. 
This fact also supports that $Z_{\text{E2}}^{\text{a}}$ is dominated by the octupole contribution. 

One of the reasons we could obtain this elegant result is that the scattering geometry for $H\parallel [\bar{1}\ 1\ 0]$ provides an ideal situation. 
Using the wave-functions obtained from a realistic mean-field model~\cite{Sera99}, we can calculate the $Z$-factors for each rank and polarization channel. 
Firstly, all the $Z$-factors for the $\pi\pi'$ channel vanishes except for $Z_{\text{E2},\pi\pi'}^{(4)}$, making the measurement and analysis straightforward. 
Vanishing of the signal for the $\pi\pi'$ channel was checked by a polarization analysis using a Mo-(200) crystal analyzer at $H$=+2 T, though the data in this paper were taken without analyzing the polarization. 
Secondly, $Z_{\text{E1},\pi\sigma'}^{(2)}$ and $Z_{\text{E2},\pi\sigma'}^{(3)}$ take their maximum at $H\parallel [\bar{1}\ 1\ 0]$, giving rise to the strongest interference. 

To summarize, we have demonstrated that even and odd rank multipoles can be extracted effectively by measuring the asymmetrical intensity of RXD with respect to the field reversal, originating from the interference between the E1 and E2 resonances.  
In the present case for CeB$_6$ in $H\parallel [\bar{1}\ 1\ 0]$, this method was quite effective to extract the field dependeces of AFQ and AFO moments. 
The result for octupole showed a good agreement with that of THF at the Boron site deduced from NMR. 
Our observation directly shows that the octupole moment is, indeed, induced in the $4f$ orbital itself as well as the quadrupole moment, 
providing an evidence for the theory of field-induced multipoles in CeB$_6$. 
We expect that the field-reversal method used in the present study can be widely applied to other multipole ordering systems such as NpO$_2$, Ce$_x$La$_{1-x}$B$_6$, SmRu$_4$P$_{12}$~\cite{Yoshizawa08}, and so on. 

The authors wish to acknowledge R. Shiina, T. Nagao and K. Hanzawa for valuable discussions. 
This work was supported by a Grant-in-Aid for Scientific Research (No. 16076202) from MEXT, Japan. 
The synchrotron experiments were performed under the approval of the Photon Factory Program Advisory Committee (No. 2005S2-003 and 2008S2-004).


\end{document}